\documentclass[aps,preprint,amsmath,amssymb,nofootinbib]{revtex4}
\usepackage{graphicx,color}

\def\be{\begin{eqnarray}}
\def\ed{\end{eqnarray}}
\def\ga{\gamma}
\def\non{\nonumber}
\def\hmp{\hat m_{K}}
\begin{document}

\title{\Large \bf Leptoquark on $P\to \ell^{+} \nu$, FCNC and LFV}
\date{\today}

\author{ \bf  Rachid Benbrik$^{1}$\footnote{Email:
rbenbrik@phys.cycu.edu.tw}  and Chuan-Hung
Chen$^{2,3}$\footnote{Email: physchen@mail.ncku.edu.tw}
 }

\affiliation{
$^{1}$Department of Physics, Chung Yuan Christian University, Chung-Li 320, Taiwan  \\
 $^{2}$Department of Physics, National
Cheng-Kung
University, Tainan 701, Taiwan \\
$^{3}$National Center for Theoretical Sciences, Hsinchu 300, Taiwan
}

\begin{abstract}
Motivated by the disagreement between the experimental data and
lattice calculations on the decay constant of the $D_{s}$ meson, we
investigate leptoquark (LQ) contributions to the purely leptonic
decays of  a pseudoscalar (P). We concentrate on the LQs which only
couple to the second-generation quarks before the electroweak
symmetry breaking and  we discuss in detail how flavor symmetry
breaking effects are brought into the extension of the standard
model after the spontaneous symmetry breaking. We show that the
assumption of the hermiticity for the fermion mass matrices can not
only  reproduce the correct Cabibbo-Kobayashi-Maskawa and
Maki-Nakagawa-Sakata matrices, but also  reduce the number of
independent flavor mixing matrices and lead to $V^{R}_{f}=V^{L}_{f}$
with $L(R)$ denoting the chirality of  the f-type fermion.
Accordingly, it is found that the decays $D_{s,d}\to \ell^{+} \nu$,
$B^{+}\to \tau^{+} \nu$ and $B_c\to \ell^{+} \nu$ have a strong
correlation in parameters. We predict that the decay constant of the
$B_c$ meson calculated by the lattice could be less than the
experimental data by $23\%$. Intriguingly, the resultant upper
limits of branching ratios for $D\to \mu^{+} \mu^{-}$ and $\tau\to
\mu (\pi^0,\, \eta,\, \eta',\, \rho,\, \omega)$ are found to be
around $ 5.1\times 10^{-7}$ and $(2.6,\, 1.5,\, 0.6,\, 7.4,\,
4.8)\times 10^{-8}$, which  are below and close to the current
experimental upper bounds, respectively.

\end{abstract}
\maketitle

As many puzzles such as matter-antimatter
asymmetry, neutrino oscillations and dark matter etc are unsolved,
it is clear that the standard model (SM) only describes parts of
the universe and
should be regarded as an effective theory at the electroweak scale. To
explore
the unknown parts, searching for
new physics
effects that do not belong to the SM becomes very important.
However, since most
measurements are eventually
resulting from
the SM, by naive speculation, the new effects should be small
and difficult to be found
out. Therefore, where we can uncover the
new physics should be addressed in the first stage to look for new
physics. Usually, the rare decays
with
 the suppressed SM contributions
 are considered to be the good candidates. In addition,
through precision measurements, finding a sizable deviation from the
theoretical expectation
provides another direction to search
for the new effects.

Recently, via the observations of $D_s\to \ell^+ \nu$ decays, CLEO
\cite{CLEO1,CLEO2} and BELLE \cite{BELLEDs} collaborations have
measured the decay constant of $D_{s}$ to be
 \be
f_{D_s} &=& 274 \pm 10 \pm 5\  \textrm{MeV}\ \ \ (\textrm {CLEO}) \,,\non \\
f_{D_s} &=& 275 \pm 16 \pm 12\  \textrm{MeV} \ \ \ (\textrm{BELLE})
\,,
\label{Data1}
 \ed
where the result by CLEO is the average of  $\mu^+ \nu$ and $\tau^+
\nu$ modes, while the BELLE's one is only from $\mu^+ \nu$. By
combining the radiative corrections from $D_s\to \gamma \ell \nu$,
the average of the two data in Eq. (\ref{Data1}) is
 \cite{RS}
 \be
 f_{D_s}= 273 \pm 10\  \textrm{MeV}\,. \label{eq:fdsexp}
  \ed
More information on the measurement from other experiments
can be found in Ref.~\cite{RS}. Furthermore, if we compare the measured
value with the recent lattice calculation \cite{UnQQCD},
 given by
 \be
f_{D_s} &=& 241 \pm 3\  \textrm{MeV}\ \ \ (\textrm {HPQCD+UKQCD})\,,
\label{eq:fdslatt}
%
 \ed
we see clearly that $3.2$ standard deviations from data have been
revealed in the purely leptonic $D_s$ decays \cite{RS,Stone}. That is, a
correction of $10\%$ to $f_{D_s}$ is needed. Does the discrepancy
indicate new physics or the defeat of the  theory? Although the answer to
the question is not conclusive yet,
by following the new
CLEO's result on the decay constant of $D^+$
\cite{CLEOD}:
 \be
f_{D}=205.8 \pm 8.5 \pm 2.5 \textrm{MeV} \ \ \ (\textrm{CLEO})\,,
 \ed
which is in a
 good agreement with the lattice calculation
\cite{UnQQCD}:
 \be
f_{D}=208 \pm 4 \textrm{MeV} \ \ \ (\textrm{Lattice}),
 \ed
it seems to tell us that the lattice improvement may not be large
enough to singly compensate the quantity that is more than $3$
standard deviations in $f_{D_s}$.

Inspired by the above interesting measurements, the authors in
Ref.~\cite{DK} propose that new interactions associated with
leptoquarks (LQs) might resolve the anomaly of $f_{D_s}$. However,
the assumption adopted by Ref.~\cite{DK} that the LQs only couple to
the second-generation quarks seems to be oversimplified. It has been
known that up-type and down-type quark mass matrices can not be
diagonalized simultaneously. Therefore, if the LQ couples to up- and
down-type quarks at the same time, after the spontaneous symmetry
breaking (SSB), the flavor mixing matrices to diagonalize the
quark mass matrices will be introduced so that intergenerational
couplings in quarks become inevitably \cite{Leurer}. To generalize
the approach of Ref.~\cite{DK}, in this paper, besides we discuss
how flavor mixing effects influence the decays $P\to \ell^{+} \nu$
and how the number of free parameters can be diminished, we also
investigate the implications of LQ interactions on the processes
with flavor changing neutral current (FCNC) and lepton flavor
violation (LFV). We note that the effects of the charged Higgs with
a large $\rm\tan\beta$ in the ordinary two-Higgs-doublet models are
destructive contributions to the SM \cite{RS,ACPRD75}, more
complicated multi-Higgs doublets are needed to get the enhancement
\cite{DK}. In addition, other models such as R-parity violation in
supersymmetric models might also provide the solution
\cite{Rparity}.
However, due to the parameters in different quark
flavors having no correlation, the models have a less predictive
power.

In order to examine
the effects of a light LQ in a systematic way,
the LQ model is built based on the gauge symmetries $SU(3)_C\times
SU(2)_L\times U(1)_Y$.
To simply display the role of the LQ on the low
energy leptonic decays, FCNC and LFV, the LQ in this paper is
limited to the $SU(2)_L$ singlet $S_1$ with the  charge of -$1/3$.
To avoid the proton decays, the LQ does not couple to diquarks.
Indicated by the inconsistent results
in the $D_s$ leptonic decays, we
consider that before the SSB, the LQ only couples to the
second-generation quarks and the interactions in the weak
eigenstates are given as \cite{DK, benbrik}
 \be
{\cal L}_{LQ} &=& \left(  \bar E g_{L} i\tau_2 P_R Q^{c}_{2}
+  \bar {\ell} g_{R} P_L c^{c}  \right )S_{1} +H.c.\,, \non\\
&=&  \left( \bar \ell  g_{L} P_R c^{c} - \bar \nu_{\ell} g_{L} P_{R}
s^{c} \right) S_1 +  \bar\ell  g_{R} P_L c^{c} S_{1} +H.c.\,,
\label{eq:lq}
 \ed
where $g_{L(R)}$ denotes a 3-component effective coupling and is
represented by $g^{T}_{\alpha}=(g_{\alpha e}, g_{\alpha \mu},
g_{\alpha \tau})$ with $\alpha=L$ and $R$, $Q^{T}_{2}=( c, s)$,
$f^{c}= C\ga^{0} f^* = C\bar
f^{T}$
 ($C=i\ga^{2}\ga^{0}$) describes the anti-fermionic state, $\tau_2$ is the 2nd Pauli matrix,
$E^{T}=(\nu_{\ell}, \ell)$ with $\ell=e,\, \mu,\, \tau$, and
$P_{L(R)}=(1\mp \ga_{5})/2$. Since the flavor mixing effects are
governed by the Yukawa sector, we write the sector as
 \be
 {\cal L}_{Y} &=& -\bar Q_L Y_{U} U_{R} \tilde H - \bar Q_{L} Y_D
 D_R H - \bar L Y_{L} \ell_{R} H + H.c. \label{eq:Yu}
 \ed
where $H$ is the SM Higgs doublet and $\tilde{H}=i\tau_{2} H^*$.
Implicitly, the flavor indices are suppressed. In addition, it is
known that the flavor changing effects in the SM only appear in processes
related to the
charged weak currents,
while the weak interactions in
weak eigenstates are expressed by
 \be
{\cal L}_{W} &=& -\frac{g}{\sqrt{2}}\left( \bar U_{L} \ga_{\mu}
D_{L} + \bar N_{L} \ga_{\mu} E_{L}\right) W^{+\mu} + H.c.
\label{eq:W}
 \ed
with $g$ being the gauge coupling of $SU(2)_{L}$. After introducing
the relevant pieces, in the following we discuss after the SSB how the
flavor mixing effects are brought into the effective interactions
and how they can be controlled through the notable patterns of mass
matrices.

It has been known
 that Eq.~(\ref{eq:Yu}) has $SU(3)_Q\times
SU(3)_{D}\times SU(3)_{U}$ \cite{MM} flavor symmetries.
As
the new
LQ interactions break the flavor symmetries, we
have to be more careful to choose the convention. The one used in
the SM is not suitable anymore for the new interacting terms. After
the
SSB, the masses of fermions are obtained by diagonalizing the Yukawa
matrices denoted by $Y_f$ with $f=U,\, D,\, E,\, N$. Although we do
not display the mass matrix for neutrinos,
due to the
observations of the neutrino oscillation, we consider that neutrinos are
massive particles. We will show that the induced effects such as
the
Maki-Nakagawa-Sakata (MNS) matrix \cite{MNS} do not explicitly
emerge after summing up the three neutrino species.
To diagonalize the mass matrices of fermions, we introduce the
unitary matrices through
 \be
 f^{p}_{\alpha} & =& V^{\alpha}_{f} f^{w}_{\alpha}\,,
 \ed
where $p(w)$ represents the physical (weak) state and $\alpha$
denotes the left or right-handness. Straightforwardly,
Eq.~(\ref{eq:W}) becomes
 \be
 {\cal L}_{W} &=& -\frac{g}{\sqrt{2}}\left( \bar u_{L} V_{CKM} \ga_{\mu}
 d_{L} + \bar \nu_{L} V_{MNS} \ga_{\mu} \ell_{L}\right) W^{+\mu} + H.c.
 \label{eq:W2}
 \ed
Here, $V_{CKM}= V^{L}_{U} V^{L\dagger}_{D}$ and $V_{MNS}= V^{L}_{N}
V^{L\dagger}_{E}$ stand for the Cabibbo-Kobayashi-Maskawa (CKM)
\cite{CKM} and MNS matrices, respectively. Clearly, besides the CKM
matrix, if we regard the neutrinos as massive particles, we bring in
a new mixing matrix for leptons. However, does $V_{MNS}$ have any
effects on the low energy leptonic decays? The answer to the
question in the SM is obvious. Since  the neutrinos in hadronic decays
are regarded as missing particles and are not detected, when one
calculates the decay rate for the process, it is needed to sum up
all neutrino species and the squared amplitude is associated with
$ \sum _{\nu} (V_{MNS})_{\nu \ell}(V^{\dagger}_{MNS})_{\ell \nu}$.
With the unitarity property,  the result becomes $ \sum _{\nu}
(V^{\dagger}_{MNS})_{\ell \nu}(V_{MNS})_{\nu \ell} =1$ so that the
effects of $V_{MNS}$ do not show up explicitly. In sum,
$V_{MNS}$ in Eq.~(\ref{eq:W2}) could be rotated away by redefining
the neutrino fields, {\it i.e.} the
neutrinos
 produced via weak interactions
  are not the mass
states
propagating in the vacuum. Will the nonrotated
$V_{MNS}$ appear in the LQ interactions?
To
 answer the question,
we need to do more analysis on the LQ sector.

With the introduced unitary matrices, Eq.~(\ref{eq:lq}) in terms of
physical states is transformed as
 \be
{\cal L}_{LQ} &=& \bar\ell_{L} \tilde g_{L} V^{L^T}_{U2} u^{c}_{L}
S_{1} - \bar \nu V_{MNS} \tilde g_{L} V^{L^T}_{D2}  d^{c}_{L} S_{1}
 \non\\
&+& \bar\ell_{R} \tilde g_{R} V^{R^T}_{U2}  u^{c}_{R} S_{1}+H.c.
\label{eq:lq2}
 \ed
where $V^{\alpha}_{U2}$, $V^{L}_{D2}$ and $\tilde g_{\alpha}$ are
3-component columns,
represented by $V^{\alpha^T}_{U2} = \left(
V^{\alpha}_{U12},\, V^{\alpha}_{U22},\, V^{\alpha}_{U32} \right)$,
$V^{L^T}_{D2} = \left( V^{L}_{D12},\, V^{L}_{D22},\, V^{L}_{D32}
\right)$ with $V^{L}_{D}= V^{\dagger}_{CKM} V^{L}_{U}$ and
$\tilde{g_{\alpha}} = V^{L^{\dagger}}_{\ell} g_{\alpha}$, respectively. Clearly,
we see that
$V_{MNS}$ appears in Eq.~(\ref{eq:lq2}).
Nevertheless, like the SM, the explicit form of $V_{MNS}$ can be rotated away by
transforming the physical neutrino states to flavor states.
Meanwhile, unlike the case in the SM where
$\nu_{\ell}$
in a process is always associated with the corresponding charged
lepton $\ell$, in the LQ model, for each charged lepton inevitably
we have to consider all possible neutrino flavors.

Using Eqs.~(\ref{eq:W2}) and (\ref{eq:lq2}) with removing $V_{MNS}$,
the effective Hamiltonian for $P\to \ell^+ \nu$ related decays are
found to be
 \be
{\cal H}( u_k \to d_i \ell^{+}_{\ell} \nu_{j}) &=&
\frac{4G_F}{\sqrt{2}}( V^{\dagger})_{ik} \delta_{\ell j} \bar  d_{i}
\ga_{\mu} P_L u_{k} \bar \nu_{j}  \ga^{\mu} P_L \ell_{\ell} \non
\\
&+&\frac{(C^{L^\dagger}_{dv})_{ji} (C^{L}_{ul})_{k\ell} }{2m^2_{LQ}}
\bar  d_{i} \ga_{\mu} P_L u_{k}
\bar \nu_{j}  \ga^{\mu} P_L \ell_{\ell} \non\\
&-& \frac{(C^{L^\dagger}_{dv})_{ji} (C^{R}_{ul})_{k\ell}
}{2m^2_{LQ}} \bar d_{i}  P_R u_{k}
\bar \nu_{j}  P_R \ell_{\ell} \non\\
&+&  \frac{(C^{L^\dagger}_{dv})_{ji} (C^{R}_{ul})_{k\ell}
}{16m^2_{LQ}} \bar d_{i}  \sigma_{\mu \nu}  u_{k}  \bar \nu_{j}
\sigma^{\mu\nu} P_R \ell_{\ell} + H.c. \label{eq:Ham}
 \ed
where
we have used $V$
as
$V_{CKM}$,
the indices $i,\, j,\, k$ and $\ell$ denote the possible flavors,
 \be
C^{L}_{ul} &=& V^{L^*}_{U2} \tilde{g}^{\dagger}_{L}\,, \ \ \
C^{R}_{ul} = V^{R^*}_{U2} \tilde{g}^{\dagger}_{R},\ \ \ C^{L}_{dv} =
V^{L^*}_{D2} \tilde{g}^{\dagger}_{L} \label{eq:rate}
 \ed
are $3\times 3$ matrices, and $\sigma_{\mu\nu}=i[\ga_{\mu},
\ga_{\nu}]/2$. Although tensor-type interactions could be generated
in Eq.~(\ref{eq:Ham}), since they cannot contribute to two-body
leptonic decays, hereafter we will not  discuss them further.
Therefore, there are two main types of four-fermion operators in
Eq.~(\ref{eq:Ham}), one is $(V-A)\times (V-A)$, which is the same as
the SM, and the other is $(S\pm P)\times (S\pm P)$. For $P\to
\ell^{+} \nu$ decays, we will see that the former will lead to the
helicity suppression, whereas
 the latter does not. On the contrary, for
$D\to \bar K \ell^{+} \nu$ decays where the lattice calculations
have been consistent with the experimental data, the latter has
the helicity suppression whereas
the former does not. Consequently, $D\to \bar K \ell^{+} \nu$ will
directly give strict constraints on the parameters
$\tilde{g}_{L\ell}$.
Since the new physics effects are considered perturbatively, if we
only keep the leading effects and neglect the higher order in
$\tilde{g_{\alpha}}$, the partial decay rate for $P\to \ell^{+} \nu$
is found to be
 \be
\Gamma(P\to \ell^{+} \nu) &=& \Gamma^{SM}(1+X^{UD}_{\ell}+Y^{UD}_{\ell})\,,\non\\
X^{UD}_{\ell}&\approx & \frac{\sqrt{2}}{4G_Fm^2_{LQ}} \frac{1}{
|V_{UD}|^2}Re\left[V^{*}_{UD}
(C^{L^\dagger}_{ul})_{\ell U}(C^{L}_{dv})_{ D\ell}\right]\,, \non \\
Y^{UD}_{\ell} &\approx & \frac{\sqrt{2}}{4G_Fm^2_{LQ}}
\frac{m^0_{P}}{m_{\ell} |V_{UD}|^2}Re\left[V^{*}_{UD}
(C^{R^\dagger}_{ul})_{\ell U}(C^{L}_{dv})_{ D\ell} \right]
\label{eq:Xlq}
 \ed
with $m^0_{P}=m^2_P/(m_U + m_D)$ and
 \be
\Gamma^{SM}&=&\frac{G^2_F}{8\pi} \left| V_{UD} \right|^2f^2_{P}
m^2_{\ell} m_{P} \left(1-\frac{m^2_{\ell}}{m^{2}_{P}} \right)^2\,,
\label{eq:brSM}
 \ed
where the decay constant $f_P$ is defined by
 \be
\langle 0| \bar D \ga_{\mu} \ga_5 U | P(p) \rangle &=& i
f_{P}p_{\mu }\,,\non\\
\langle 0| \bar D  \ga_5 U | P(p) \rangle &=& -i
f_{P}\frac{m^2_{P}}{m_D+m_U}\,. \label{eq:DC}
 \ed
 Since CP problem is not concerned in this paper, for
a further simplification of our numerical analysis, the weak phases will be
tuned to
zero. Then, $X^{UD}_{\ell}$ and $Y^{UD}_{\ell}$ can be
shortened as
 \be
  X^{UD}_{\ell}& \approx & \frac{\sqrt{2}}{4G_Fm^2_{LQ}} \frac{1}{
V_{UD}}(C^{L^\dagger}_{ul})_{\ell U}(C^{L}_{dv})_{D\ell }\,,\non\\
 Y^{UD}_{\ell}& \approx & \frac{\sqrt{2}}{4G_Fm^2_{LQ}} \frac{m^{0}_{P}}{m_{\ell}
V_{UD}}(C^{R^\dagger}_{ul})_{\ell U}(C^{L}_{dv})_{D\ell }\,.
\label{eq:X}
 \ed
Clearly, $X^{UD}_{\ell}$  and $Y^{UD}_{\ell}$ are associated with
$|\tilde{g}_{L\ell}|^2$ and $\tilde{g}^*_{L\ell} \tilde{g}_{R\ell}$,
respectively. We note that the capital symbol of $U(D)$ denotes the
up (down)-type quark in a specific decay. For instance,
$X^{cs}_{\ell}$ and
 $X^{ud}_{\ell}$ are for $D_s$ and
$\pi$ decays, respectively.

Before doing the numerical analysis, we need to know how many free
parameters are involved in the model and how to reduce the number of
parameters. From Eq.~(\ref{eq:lq}), it is obvious that six free
parameters
from $g_L$ and
$g_R$ are introduced in the original LQ model. These
parameters are associated with the flavors of the charged leptons. After the
SSB, due to the misalignment between mass and interaction states, we
have to bring the new unitary matrices $V^{\alpha}_{f}$ to
diagonalize the Yukawa matrices. Except that $V_{CKM}= V^{L}_{U}
V^{L^\dagger}_{D}$ is known by fitting the data, the
elements in the unitary matrices are
usually  regarded as free
parameters.
In general, there is no any relationship among the flavor mixing
matrices. Nevertheless, by utilizing the
experimental data, we can obtain some clues
to sense the information on
mixing matrices. It is known that the determination of the flavor mixing
matrices $V^{\alpha}_{f}$ is governed by the detailed patterns of
the mass matrices. According to
 $V_{CKM}$ being approximately
a
unity matrix, people have found that  the quark mass matrices are
very likely aligned and have the relationship of ${\cal M}_{D}={\cal
M}_{U} + \Delta(\lambda^2)$ with ${\cal M}_{U(D)}=M_{U(D)}/m_{t(b)}$
\cite{qm1,qm2,qm3}. In other words, the structure of
$V^{\alpha}_{U}$
 should be similar to $V^{\alpha}_{D}$.
Furthermore, it has been shown that a simple pattern of the mass
matrix, proposed by Ref.~\cite{mass} with
 \be
 M_{f}&=&P^{\dagger}_{f} \bar M_{f} P_{f}\
{ \rm with}\ \bar M_{f}=\left(
        \begin{array}{ccc}
          0 & A_{f} & 0 \\
          A_{f} & D_{f} & B_{f} \\
          0 & B_{f} & C_{f}  \\
        \end{array}
      \right)\,, \non\\
      P_{f} &=& (e^{i\theta_{f1}},\, e^{i\theta_{f2}},\, e^{i\theta_{f3}})\,,
       \label{eq:mass}
 \ed
could lead to reasonable structures for the mixing angles and CP
violating phase in the CKM matrix just in terms of the quark masses.
Using the current accuracy of data, the mass patterns of
Eq.~(\ref{eq:mass}) have been reanalyzed and applied to lepton
masses by the authors in Ref.~\cite{MN}. It is found that
 the elements of $V_{CKM}$ can satisfy with current accuracy of data
and the  component of $(V_{MNS})_{13}$ can be consistent with
present experimental constraint as well. Although the phenomenological
patterns may not be the general form, due to the support of
experiments, the resultant flavor mixing matrices could be taken as
a clue to
the true mass matrices.

Inspired by the fascinating mass matrices and their results, we
speculate that
to avoid the restricted patterns shown in
Eq.~(\ref{eq:mass}), the mass matrices could be extended to those
which  not only
 own the main character of Eq.~(\ref{eq:mass})
but also provide the relationship between $V^{R}_{f}$ and
$V^{L}_{f}$. Accordingly, we find that the criterion to get a more
general property of Eq.~(\ref{eq:mass}) could be established if the
mass matrices are hermitian. It is worth mentioning that the
hermitian mass matrices could be naturally realized in gauge models
such as left-right symmetric models \cite{hermitian_y}. Furthermore,
the hermiticity is helpful to solve the CP problem in models with
supersymmetry (SUSY) \cite{ABKL}, which
 has an important implication
on  CP violation in Hyperon decays \cite{Chen_PLB521}. With the
hermiticity, we
obtain the results $V^{L}_{f}=V^{R}_{f}\equiv
V_{f}$. Via $V_{U} = V V_{D}$
from the definition of
the CKM matrix, intriguingly the number of independent flavor mixing
matrices in the quark sector could be reduced to one and
the unknown flavor mixing matrix is chosen to be
$V_{D}$ for our
following analysis.

After setting up the model and the associated parameters,
subsequently we study the constraints on the free parameters and
their relative implications. Firstly, we discuss the limits of $D\to
\bar K \ell^+ \nu$. As mentioned early, the effects of
$\tilde{g}_{R}$ for $D\to \bar K  \ell^{+} \nu$ are helicity
suppressed.
Here we only display the constraints on
$\tilde{g}_{L}$. By the effective Hamiltonian of Eq.~(\ref{eq:Ham}),
the transition matrix element for $D\to \bar K \ell^{+} \nu $ can be
written as
 \be
{\cal M}(D\to \bar K \ell^{+} \nu)_{SM+LQ} &=& -\frac{G_F}{\sqrt{2}}
V^*_{cs}\sum_{j} \left( \delta_{j\ell} + \frac{\sqrt{2}}{8G_F
m^2_{LQ}} \frac{(C^{L^\dagger}_{ul})_{\ell c}
(C^{L}_{dv})_{sj}}{V_{cs}} \right) \non\\
&\times& \langle K| \bar s \ga_{\mu} c | D \rangle \bar \nu_{j}
\ga^{\mu}(1-\ga_5) \ell_{\ell}\,,
 \ed
where the sum is to include all neutrino species  and the $D\to
\bar K$ form factors can be parametrized by
  \be
 \langle  \bar K| \bar s \ga_{\mu} c |  D \rangle &=& f_{+}(q^2) \left(
P_{\mu}-  \frac{P\cdot q}{q^2} q_{\mu} \right) + f_{0}(q^2)
\frac{P\cdot q}{q^2}q_{\mu}\,.
 \ed
If the effects of the 2nd order in $\tilde{g}_{L}$ are neglected, a
simple expression for $D\to \bar K \ell^{+} \nu$ is given by
 \be
 {\cal B}(D\to \bar K \ell^{+} \nu)_{Exp} &=& \left( 1+
 X^{cs}_{\ell}\right) {\cal B}(D\to K \bar \ell \nu)_{SM}\,.
 \ed
With $V_{U}=V V_{D}$, the effective coupling
$(C^{L^\dagger}_{ul})_{\ell c} (C^{L}_{dv})_{s\ell}$ could be
expressed as
 \be
(C^{L^\dagger}_{ul})_{\ell c} (C^{L}_{dv})_{s\ell} &=& \left(V_{cd}
V_{D12} + V_{cs} V_{D22} + V_{cb} V_{D32} \right)
V^{*}_{D22}|\tilde{g}|^2_{L\ell}\,.
 \ed
Since the off-diagonal elements of $V_{D}$ represent the flavor
symmetry breaking effects, according to
Eq.~(\ref{eq:mass}),  the diagonal elements of $V_{D}$ are roughly
order of unity while the off-diagonal elements are order of
$\sqrt{m_i /m_j}$ with $j>i$ \cite{qm3,mass,CG_PLB661}. As a result,
 $X^{cs}_{\ell}$ could be written as
  \be
 X^{cs}_{\ell}& \approx & 2\frac{m^2_W}{m^2_{LQ}} \left\{
\begin{tabular}{c}
  {$|\tilde{g}_{Le}|^2/g^2$ \ \ \rm for $\ell= e$}\,,\\
  {$|\tilde{g}_{L\mu}|^2/g^2$ \ \ \rm for $\ell=\mu$}
\end{tabular} \right.
 \ed
where $G_F/\sqrt{2}= g^2/8m^2_{W}$ has been used.
 From the
data~\cite{PDG06} and the recent unquenched lattice
calculation~\cite{Lattice},
 given by
 \be
 \Gamma(D^0\to K^- \ell^{+} \nu)_{Exp} &=& (8.17 \pm 0.48)\times 10^{-2}
 ps^{-1}\hspace{2cm}\
 (\textrm{PDG})\,,\non\\
\Gamma(D^0\to K^- \ell^{+} \nu)_{Latt} &=& ( 9.2 \pm 0.7 \pm 1.8 \pm
0.2 )\times 10^{-2} ps^{-1}\ \ \
 ( \textrm{Lattice} )\,,
 \ed
obviously the theoretical calculation is consistent with the
experimental value, {\it i.e.} we can set $\tilde{g}_{L\ell}$
as
small as possible. In order to sense the order of magnitude of the
parameters, we require that new physics effects are only less than
$1\sigma_{Exp}$, {\it i.e.}
 \be
 \frac{\Gamma_{Exp} - \Gamma_{Latt}}{\Gamma_{Latt}} = X^{cs}_{\ell} <
 8\%\,.
 \ed
Using $g\approx 0.67$ and $m_W\approx 80$ GeV, we get
 \be
\left( \frac{\tilde{g}_{L\ell^{\prime}}}{m_{LQ}} \right)^2<
2.8\times 10^{-6} \textrm{\ GeV}^{-2} \label{eq:limit_gl}
 \ed
with $\ell^{\prime}=e,\, \mu$.



Now, we study the LQ effects on $D_{s}\to \ell^{+} \nu$ decays where
the disagreement between theory and experiment shows up.
In terms of
the previous analysis, although the LQ only couples to the
second-generation quarks, through the flavor mixing matrices,
the LQ could also couple to the quarks of the first and
third generations. Therefore, besides $D_s \to \ell^{+} \nu$ decays,
we can also study the
processes $D_d\to \ell^{+} \nu$ and
$B_u\to \tau^+ \nu$, in which the involving parameters are
correlated each other. Taking $V_{cs}\approx 1$, $V_{cd}=-\lambda\simeq 0.22$,
$V_{D22}\approx 1$
and neglecting the subleading terms, from Eq.~(\ref{eq:X}) the
effects of LQ to $D_s\to \ell^{+} \nu$, $D_{d}\to \ell'^{+} \nu$ and
$B_u \to \tau^+ \nu$ can be  simplified to be
 \be
X^{cs}_{\ell} &\approx & 2\frac{m^2_{W}}{ m^2_{LQ}}
\frac{\tilde{g}^2_{L\ell}}{g^2} \,,\ \ \  Y^{cs}_{\ell} \approx
2\frac{m^2_{W}}{ m^2_{LQ}} \frac{m^0_{D_s}}{m_{\ell}}
\frac{\tilde{g}^*_{L\ell} \tilde{g}_{R\ell}}{g^2}\,,\non\\
X^{cd}_{\ell'} &\approx & \frac{V^*_{D12}}{-\lambda}
X^{cs}_{\ell'}\,,\ \ \ Y^{cd}_{\ell'} \approx
\frac{V^*_{D12}}{-\lambda} \frac{
m^0_{D_d}}{m^0_{D_s}} Y^{cs}_{\ell'}\,,\non\\
X^{ub}_{\tau} &\approx & \left( V_{D12}+\lambda
\right)\frac{V^*_{D32}}{ V_{ub}} X^{cs}_{\tau}\,,\ \ \ Y^{ub}_{\tau}
\approx  \left( V_{D12}+\lambda  \right)\frac{V^*_{D32}}{ V_{ub}}
\frac{m^{0}_{B}}{m^0_{D_s}}Y^{cs}_{\tau}\,, \label{eq:XY}
 \ed
 respectively.
Clearly, the parameters contributing to $D_{s}\to \ell^{+} \nu$ will
also affect the decays $D_{d}\to \ell'^{+} \nu$ and $B_u\to \tau^+
\nu$. Moreover, since the decay rate for $P\to \ell^{+} \nu$ is
directly related to the decay constant of the P-meson,
to display
the new physics effects, we express the connection of the observed decay
constant with the lattice calculation to be
 \be
 f^{Exp}_{P} = f^{Latt}_{P} \sqrt{ 1 + X^{UD}_{\ell} +
 Y^{UD}_{\ell}}\approx  f^{Latt}_{P} \left(1+ \frac{X^{UD}_{\ell}+ Y^{UD}_{\ell}}{2} \right)
 \,.
 \ed
To  explain the anomalous results occurred in $D_{s}\to (\mu^{+},\,
\tau^{+}) \nu$ shown in Eqs.~(\ref{eq:fdsexp}) and
(\ref{eq:fdslatt}), the new physics at least should enhance
$f_{D_s}$ by $10\%$, that is, $X^{cs}_{\mu(\tau)}+
Y^{cs}_{\mu(\tau)}$ should be around $20\%$. Due to $X^{cs}_{\ell} <
8\%$, we see that the dominant contributions are from
$Y^{cs}_{\ell}$. For simplicity, we will ignore the effects of
$X^{cs}_{\ell}$ and adopt $Y^{cs}_{\ell}\approx 0.2$.

For $Y^{cd}_{\ell'}$, now we have the ambiguity in sign of
$V_{D12}$, denoted by Sign$[V_{D12}]$. To understand the sign, we
can refer to the result of Eq.~(\ref{eq:mass}) in which
Sign$[V_{D12}]<0$ \cite{MN}. With $Y^{cs}_{\ell'}=0.2$, we get
$Y^{cd}_{\ell'}=0.18|V_{D12}|/\lambda$. Since the results of the
data and the lattice result on $f_{D}$ are consistent each other, to
fit the data within $1\sigma$, one can find that the value of
$|V_{D12}|$ should be less than $0.57\lambda$ where if $f^{\rm
Latt}_{D}=204$ MeV is used,
which leads to $f_{D}\approx 214.7$ MeV. As for $Y^{ub}_{\tau}$,
Sign$[V_{D32}]$ is also ambiguous. Again, the sign could be chosen
to be the same as that provided by Eq.~(\ref{eq:mass}) in which
Sign$[V_{D32}]>0$. Comparing with $f_{D_s}$ and $f_{D}$, although
the error of $f_{B_u}$ calculated by the lattice \cite{HPQCD} is
somewhat larger, due to the large enhancements of $1/|V_{ub}|\sim
1/\lambda^{4}$ and $m^0_{B}/m^0_{D_s}$, $B_u \to \tau^+ \nu$ can
still give a strict limit on $V_{D32}$.  Using the averaged value of
$V_{ub}=3.9\times 10^{-3}$ \cite{RS} and $f_B=216$ MeV, the SM
prediction on the branching ratio (BR) is ${\cal B}(B_u \to \tau^+
\nu)=1.25\times 10^{-4}$. Taking the data with $1\sigma$ error and
${\cal B}(B_u\to \tau^+ \nu)= 1.85\times 10^{-4}$ as the upper
bound, we obtain $V_{D32} < 0.043$. By combining the above analysis,
the instant predictions are the BRs for $B_{c}\to  \ell^{+} \nu$
decays. Similar to Eq.~(\ref{eq:XY}), the LQ contributions to $B_c$
decays could be written as
 \be
Y^{cb}_{\ell} = \frac{V_{U22} V^*_{D32}}{V_{cb}}
\frac{m^0_{B_c}}{m^0_{D_s}} Y^{cs}_{\ell} \,.
 \ed
Adopting $V_{U22}\approx 1$, $V_{D32}\approx 0.043$, $V_{cb}\approx
0.042$ and $Y^{cs}_{\ell}\approx 0.2$, we immediately find
$Y^{cb}_{\ell} \approx 0.49$. In other words, we predict that the
calculation of the lattice on $f_{B_c}$ could have $\sim 23\%$ below
the observation of  the experiment.
According to above analysis, we see clearly that even in
the
restricted case, where the fermion mass matrices are hermitian, the
explanation of the $D_s$ puzzle in terms of the LQ remains viable
despite constraints from other flavor processes.

With the constraints on the parameters of the LQ model, in the
following we study the implications of the LQ effects on the decays
associated with FCNC and LFV. Firstly, we discuss the $D\to \mu ^{+}
\mu^{-}$ decay. It is
known that due to the stronger Glashow-Iliopoulos-Maiani (GIM)
mechanism \cite{GIM}, the short-distance contributions to $D\to
\mu^{+} \mu^{-}$ are highly suppressed in the SM \cite{BGHP_PRD52}
and long-distance effects are small \cite{BGHP_PRD66}. The decay of
$D\to \mu^{+} \mu^{-}$ is definitely a good candidate to probe the
new physics effects \cite{CGYPLB655}. According to
Eq.~(\ref{eq:lq2}), we know that the dominant effective Hamiltonian
for $c\to u \mu^{+} \mu^{-}$ is from the left-right interference
terms and can be written as
 \be
{\cal H}(c\to u \mu^{+} \mu^{-}) &=& -\frac{1}{2m^2_{LQ}} \left[
\left( C^{L}_{ul}\right)_{c\mu}
\left(C^{R}_{ul}\right)^{\dagger}_{\mu u} \bar u P_{L} c \bar\mu
P_{L} \mu \right. \non\\
&+&\left.  \left( C^{R}_{ul}\right)_{c\mu}
\left(C^{L}_{ul}\right)^{\dagger}_{\mu u} \bar u P_{R} c \bar\mu
P_{R} \mu\right]+H.c.
 \ed
By combining Eqs.~(\ref{eq:rate}), (\ref{eq:DC}), (\ref{eq:XY}) and
$V_{U}=V V_{D}$, the BR for $D\to \mu^{+} \mu^{-}$ can be simplified
to be
 \be
{\cal B}(D\to \mu^{+} \mu^{-})&=& \tau_{D}\frac{m_D}{8\pi}\left(
1-\frac{4m^2_{\mu}}{m^2_D}\right)^{1/2} \left( \frac{G_F}{\sqrt{2}}
\frac{m^0_{D}}{m^0_{D_s}}f_D m_{\mu}\right)^2 \non \\
&\times& |Y^{cs}_{\mu}|^2 |V_{D12}+\lambda|^2\,.
 \ed
Using $Y^{cs}_{\mu}\approx 0.2$ and $V_{D12}\approx -0.57\lambda$,
the values of BR with various values of $f^{\rm Latt}_{D}$ are
presented in Table \ref{tab:Dmumu}. Interestingly, the LQ
predictions satisfy and are close to the current experimental upper
bound,
given by ${\cal B}(D\to \mu^{+} \mu^{-})|_{Exp}< 5.3 \times 10^{-7}$
\cite{CDF}.

\begin{table}[hptb]
\caption{ Upper limits of the LQ on ${\cal B}(D\to \mu^{+} \mu^{-})$
with various values of $f^{\rm Latt}_{D}$. The upper bound of the
current data is $5.3 \times 10^{-7}$ \cite{CDF}. }\label{tab:Dmumu}
\begin{ruledtabular}
\begin{tabular}{cccccc}
$f^{\rm Latt}_{D}$(MeV)  & $204$ & $206$ & $208$ & $210$ & $212$
 \\ \hline
BR & $4.9\times 10^{-7}$ & $ 5.0\times 10^{-7}$ & $5.1\times
10^{-7}$ & $5.2\times 10^{-7}$ & $5.3\times 10^{-7}$
%
%
\end{tabular}
\end{ruledtabular}
\end{table}
%

The LQ interactions in Eq. (\ref{eq:lq2}) could also contribute to
the lepton flavor violating processes. Since the constraints on the
$\tilde{g}_{Re}$ are more uncertain, we only pay attention to the
decays $\tau\to \mu (P,\, V)$, in which the relevant effective
Hamiltonian is
 \be
{\cal H}_{\tau \to \mu u\bar u} &=& -\frac{1}{2m^2_{LQ}}
\left(C^{R}_{ul} \right)_{u\tau} \left(C^{R}_{ul}
\right)^{\dagger}_{\mu u} \bar u \ga^{\mu} P_{R}u \bar\mu \ga_{\mu}
P_{R} \tau +H.c.
 \ed
For the light mesons,
$u$ represents
 the up-quark. By Eq.~(\ref{eq:rate}),
the BRs for $\tau\to \mu (P,\, V)$ are given by
 \be
{\cal B}(\tau \to \mu P)&=& \tau_{\tau} \frac{f^2_P
m^3_{\tau}}{2^{10}\pi} \left( 1 -
\frac{m^2_{P}}{m^2_{\tau}}\right)^2
\frac{|\tilde{g}_{R\tau}|^2}{m^2_{LQ}}
\frac{|\tilde{g}_{R\mu}|^2}{m^2_{LQ}} |V_{D12} + \lambda|^2 \,,
\non\\
{\cal B}(\tau \to \mu V)&=& \tau_{\tau} \frac{f^2_V
m^3_{\tau}}{2^{10}\pi} \left( 1 -
\frac{m^2_{V}}{m^2_{\tau}}\right)^2 \left(
1+2\frac{m^2_{V}}{m^2_{\tau}} \right)
\frac{|\tilde{g}_{R\tau}|^2}{m^2_{LQ}}
\frac{|\tilde{g}_{R\mu}|^2}{m^2_{LQ}} |V_{D12} + \lambda|^2 \,,
\label{eq:tauPV}
 \ed
respectively.
To calculate the modes associated with $\eta$ and
$\eta^{\prime}$ mesons, we employ the quark-flavor scheme in which
$\eta$ and $\eta'$ physical states could be described by
\cite{flavor0,flavor}
\begin{eqnarray}
\left( {\begin{array}{*{20}c}
   \eta   \\
   {\eta '}  \\
\end{array}} \right) = \left( {\begin{array}{*{20}c}
   {\cos \phi } & { - \sin \phi }  \\
   {\sin \phi } & {\cos \phi }  \\
\end{array}} \right)\left( {\begin{array}{*{20}c}
   {\eta _{q} }  \\
   {\eta _{s} }  \\
\end{array}} \right) \label{eq:flavor}
\end{eqnarray}
with $\phi$ being the mixing angle, $\eta _{q}  = ( {u\bar u + d\bar
d})/\sqrt{2}$ and $\eta_{s} = s\bar s$. Accordingly, the decay
constant of $\eta^{(\prime)}$ associated with $\bar u \ga^{\mu}
\ga^{5}u$ current is given by $f_{\eta^{(\prime)}}=\cos\phi(
\sin\phi) f_{\eta_q}$. For numerical calculations, we have to know
the direct bound on the free parameter $\tilde{g}_{R\tau}/m_{LQ}$.
 From $Y^{cs}_{\ell}$ of Eq.~(\ref{eq:XY}) and the result of
Eq.~(\ref{eq:limit_gl}), the information can be obtained immediately
as
 \be
 Y^{cs}_{\ell} &\leq& 1.3\times
 10^{2} \frac{\tilde{g}_{R\ell}}{m_{LQ} m_{\ell}} \,.\non
 \ed
With $Y^{cs}_{\ell} \approx 0.2$, the direct bound on
$\tilde{g}_{R\ell}/m_{LQ}$ is found to be
 \be
\frac{\tilde{g}_{R\ell}}{m_{LQ}} \leq 1.6\times 10^{-3} m_{\ell}\,.
 \ed
By taking $\phi\approx 39^{\circ}$, $f_{\eta_q}\approx 140$ MeV
\cite{flavor}, $f_{\pi}=130$ MeV, $f_{\rho(\omega)}=216 (187)$ MeV,
$V_{D12}\approx -0.57\lambda$ and the above resultant upper limits,
the values of BRs for $\tau \to \mu (\pi^0,\, \eta,\, \eta',\,
\rho^0,\, \omega)$ decays are displayed in Table \ref{tab:tau}. We
see that interestingly the contributions of the LQ to lepton flavor
violating processes are below the current experimental upper bounds.
In addition, the predictions on the decays $\tau \to \mu (\eta,\,
\rho,\, \omega)$ are very close to the current upper bounds.
\begin{table}[hptb]
\caption{ Upper limits of BRs from the current data \cite{PDG06,BELLEtau}
and the LQ. }\label{tab:tau}
\begin{ruledtabular}
\begin{tabular}{cccccc}
Mode  & $\tau\to \mu \pi^0$ & $\tau\to \mu \eta$ & $\tau\to \mu
\eta^{\prime}$ & $\tau\to \mu \rho^0$ & $\tau\to \mu \omega$
 \\ \hline
Current limit & $1.1\times 10^{-7}$ & $ 6.5\times 10^{-8}$ &
$1.3\times 10^{-7}$ & $2.0\times 10^{-7}$ & $8.9\times 10^{-8}$ \\
This work & $2.6\times 10^{-8}$ & $1.5\times 10^{-8}$ & $0.6\times
10^{-8}$ & $7.4\times 10^{-8}$ & $4.8\times 10^{-8}$
\end{tabular}
\end{ruledtabular}
\end{table}
%

In summary, to understand the
inconsistency between the experimental data and lattice calculations
in $f_{D_s}$,
we have extended
the SM to include the LQ interactions  which involve
 only the second-generation quarks
 above the electroweak
scale. After the SSB, the
flavor mixing matrices introduced to diagonalize the mass
matrices of quarks can make the LQ couple to the first and third
generations.
We have derived
that if the mass matrices of fermions are hermitian in which the
obtained CKM and MNS matrices can be consistent with data,
besides
having $V^{R}_{f}=V^{L}_{f}\equiv V_{f}$, the independent flavor
mixing matrices are further reduced to one, say $V_{D}$.
Accordingly, it is found that the effects of the LQ on the decays
$D_{s,d}\to \ell^{+} \nu$, $B^{+}\to \tau^{+} \nu$ and $B_c\to
\ell^{+} \nu$ are correlated together. With the obtained
constraints, we predict $f^{\rm Exp}_{B_c}\approx 1.23 f^{\rm
Latt}_{B_c}$. Moreover, the upper limits of BRs for $D\to \mu^{+}
\mu^{-}$ and $\tau\to \mu (\pi^0,\, \eta,\, \eta',\, \rho,\,
\omega)$ are found to be around $ 5.1\times 10^{-7}$ and $(2.6,\,
1.5,\, 0.6,\, 7.4,\, 4.8)\times 10^{-8}$, respectively.
Interestingly, all predicted values are below and close to the
current experimental upper bounds.

\begin{acknowledgments}

This work is supported in part by the National Science Council of
R.O.C. under Grant \#s:NSC-95-2112-M-006-013-MY2 and
NSC96-2811-M-033-005.
 \end{acknowledgments}


\end{document}